\documentclass[10pt,aps,prl,superscriptaddress,amssymb,nobibnotes,reprint]{revtex4-1}

\usepackage[utf8x]{inputenc}
\usepackage{graphicx}

\newcommand{\Tm}{{T^-}}
\newcommand{\taum}{{\tau_m}}
\newcommand{\Tone}{{T_1}}
\newcommand{\gatetwo}{\mathrm{G_2}}
\newcommand{\gateone}{\mathrm{G_1}}

\begin{document}
\title{Gate-based single-shot readout of spins in silicon}
\author{A.~West}
\thanks{These authors contributed equally}
\author{B.~Hensen}
\thanks{These authors contributed equally}

\affiliation{Centre for Quantum Computation and Communication Technology,
School of Electrical Engineering and Telecommunications,
The University of New South Wales, Sydney, New South Wales 2052, Australia}
\author{A.~Jouan}
\affiliation{ARC Centre of Excellence for Engineered Quantum Systems, School of Physics, 
The University of Sydney, Sydney, NSW 2006, Australia}
\author{T.~Tanttu}
\author{C.H.~Yang}
\affiliation{Centre for Quantum Computation and Communication Technology,
School of Electrical Engineering and Telecommunications,
The University of New South Wales, Sydney, New South Wales 2052, Australia}
\author{A.~Rossi}
\affiliation{Cavendish Laboratory, University of Cambridge, J.J. Thomson Avenue, Cambridge CB3 0HE, United Kingdom}
\author{M.F.~Gonzalez-Zalba}
\affiliation{Hitachi Cambridge Laboratory, 
J.J. Thomson Avenue, Cambridge CB3 0HE, United Kingdom}
\author{F.E.~Hudson}
\author{A.~Morello}
\affiliation{Centre for Quantum Computation and Communication Technology,
School of Electrical Engineering and Telecommunications,
The University of New South Wales, Sydney, New South Wales 2052, Australia}
\author{D.J.~Reilly}
\affiliation{ARC Centre of Excellence for Engineered Quantum Systems, School of Physics, 
The University of Sydney, Sydney, NSW 2006, Australia}
\affiliation{Microsoft Corporation, Station Q Sydney, The University of Sydney, Sydney, NSW 2006, Australia}
\author{A.S.~Dzurak}
\email{b.hensen@unsw.edu.au, a.dzurak@unsw.edu.au}
\affiliation{Centre for Quantum Computation and Communication Technology,
School of Electrical Engineering and Telecommunications,
The University of New South Wales, Sydney, New South Wales 2052, Australia}

\begin{abstract}
Electron spins in silicon quantum dots provide a promising route towards realising the large number of coupled qubits required for a useful quantum processor\cite{Veldhorst2014addressable,Kawakami2014Electrical,Veldhorst2015two-qubit,Maurand2016CMOS,Zajac2016Scalable,Watson2018programmable,Zajac2018Resonantly,Huang2018Fidelity}. At present, the requisite single-shot spin qubit measurements are performed using on-chip charge sensors, capacitively coupled to the quantum dots. However, as the number of qubits is increased, this approach becomes impractical due to the footprint and complexity of the charge sensors, combined with the required proximity to the quantum dots\cite{Zajac2016Scalable}. Alternatively, the spin state can be measured directly by detecting the complex impedance of spin-dependent electron tunnelling between quantum dots\cite{Petersson2010Charge,House2015Radio,Betz2015Dispersively}. This can be achieved using radio-frequency reflectometry on a single gate electrode defining the quantum dot itself\cite{Betz2015Dispersively,Colless2013Dispersive,Gonzalez-Zalba2015Probing,Rossi2017Dispersive,ahmed_radio-frequency_2018}, significantly reducing gate count and architectural complexity, but thus far it has not been possible to achieve single-shot spin readout using this technique. Here, we detect single electron tunnelling in a double quantum dot and demonstrate that gate-based sensing can be used to read out the electron spin state in a single shot, with an average readout fidelity of 73\%. The result demonstrates a key step towards the readout of many spin qubits in parallel, using a compact gate design that will be needed for a large-scale semiconductor quantum processor.
\end{abstract}
\maketitle

Spins in silicon possess long coherence times\cite{Veldhorst2014addressable}, can couple naturally through the exchange interaction or via an engineered quantum bus, have a small qubit footprint and are amenable to mass scale semiconductor fabrication techniques\cite{Maurand2016CMOS}. Spin-dependent tunnelling to a neighbouring quantum dot\cite{Petta2005Coherent,Barthel2009Rapid} or an electron reservoir\cite{Elzerman2004Single-shot} allows qubit measurement by mapping the spin information to a detectable charge distribution. The operation of quantum point contact (QPC) and single electron transistor (SET) charge sensors to sense nearby quantum dots has led to high fidelity single-shot readout of spin based qubits\cite{Barthel2009Rapid}, a key resource for the implementation of quantum algorithms and error detection\cite{nielsen_quantum_2004,Jones2018Logical}. In order to obtain sufficient capacitive coupling, however, the sensors must be placed within a few tens of nanometres from the targeted quantum dots, meaning each sensor can only detect the charge state of a small number of quantum dots\cite{Zajac2016Scalable}. For quantum processing architectures that propose employing arrays of quantum dots\cite{Hill2015surface,Veldhorst2017Silicon,Vandersypen2017Interfacing,Jones2018Logical,Li2018crossbar}, reading all of the dots with this technique would require a high density of charge sensors, each one of which requires ohmic reservoirs and one to three gate electrodes. A gate-based readout mechanism that detects the qubit state using the same gate electrodes that define the quantum dot itself would present a significant advantage in compactness and simplicity. 

The detection of electron tunnelling using radio-frequency (rf) reflectometry has been demonstrated in a variety of quantum dot architectures\cite{Petersson2010Charge,House2015Radio,Betz2015Dispersively,Colless2013Dispersive,Gonzalez-Zalba2015Probing,Rossi2017Dispersive,ahmed_radio-frequency_2018}. By detecting shifts in the phase of the reflected signal, this technique can approach the sensitivity of state of the art charge sensors.  When combined with Pauli spin blockade in a double quantum dot, it provides direct access to the electron spin qubit information\cite{Petersson2010Charge,House2015Radio,Betz2015Dispersively,Jones2018Logical}, since electron tunnelling between two loaded dots is restricted to the spin singlet state. Gate-based sensing\cite{Colless2013Dispersive}, as opposed to employing the quantum dot source-drain electrodes\cite{Petersson2010Charge}, limits the attainable sensitivity due to the reduced geometrical capacitance, but alleviates the need for a nearby reservoir altogether. The key missing requirement for this technique to find use in a scaled architecture is the single-shot readout of a single spin.

Here we employ a silicon metal-oxide-semiconductor (SiMOS) double quantum dot architecture\cite{Veldhorst2014addressable,Veldhorst2015two-qubit,Huang2018Fidelity,Tanttu2018Controlling}, equipped with an on-chip SET to benchmark the gate-based sensing. To accumulate electrons at the silicon-oxide interface, a positive voltage must be applied on the quantum dot accumulation gates ($\gateone$,$\gatetwo$, Fig. 1a). The electrons are confined to small quantum dots under the tip of the gates by a confinement barrier gate (C) and tunnel coupled to a reservoir of electrons under gate R. Accumulation gate $\gatetwo$ is embedded in an $L-C$ resonant circuit, consisting of a surface-mount inductor $L = 400$~nH on the printed circuit board that holds the device chip, and the parasitic capacitance $C_p$, see Fig. 1c. Building on previous work in this system\cite{Rossi2017Dispersive}, we optimise the gate design for dispersive sensing by removing the possibility for electrons to accumulate under $\gateone$,$\gatetwo$ in the fan-out region of the device. Such electrons induce a gate-voltage dependent contribution to $C_p$, even if they are far away from the quantum dot\cite{Rossi2017Dispersive,croot_gate-sensing_2017} and interfere with the gate-based sensing. By extending gate C to this region, we prevent electron accumulation under $\gateone$,$\gatetwo$. Using standard reflectometry techniques, we measure the reflected amplitude and phase response (Fig. 1b), which yields a resonance frequency $f_0=266.9$~MHz and quality factor $Q = 38$, from which we determine $C_p=1/(2\pi f_0^2 L)=0.89$~pF. The description of electron tunnelling in terms of a complex impedance $Z(f)$ has been studied extensively\cite{Petersson2010Charge,House2015Radio,Colless2013Dispersive,Gonzalez-Zalba2015Probing,Cottet2011Mesoscopic}. Here we focus on the effect of inter-dot tunnelling, as this provides access to the electron spin information via Pauli spin blockade. In the case of an inter-dot tunnel coupling $t_c\gg f_0$, the effect of the electron tunnelling can be described by a quantum capacitance $C_q$, $Z(f) = \frac{1}{j 2 \pi f C_q}$, in parallel with the parasitic capacitance $C_p$. Consequently, $C_q$ causes a shift of the resonator frequency that is detected as a phase shift $\Delta \phi \approx - \pi Q C_q/C_p$, when probed at a fixed frequency $f=f_0$.

\begin{figure}[htb]
	\centering
	\includegraphics[width=80mm]{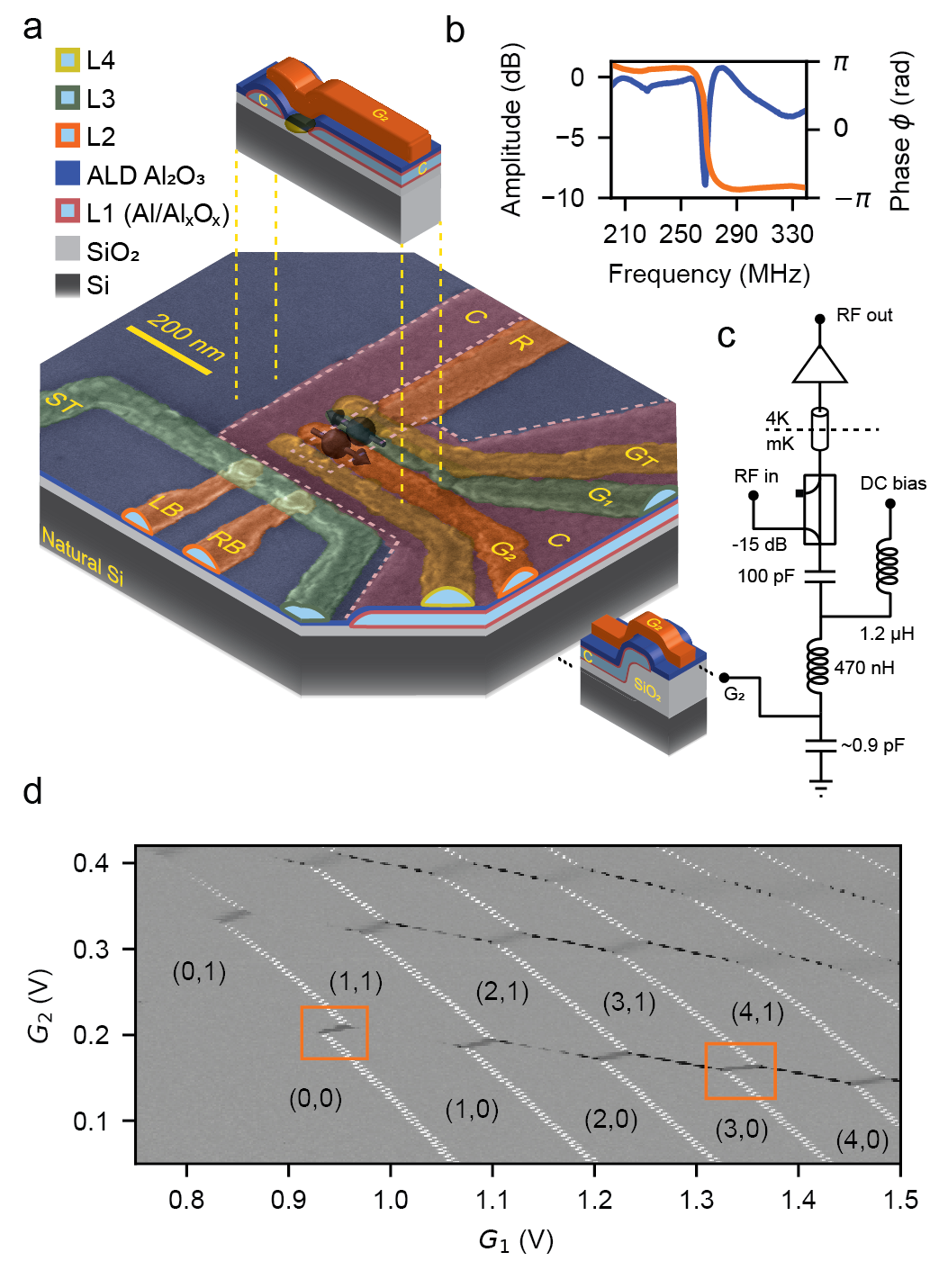}
	\caption{\label{fig:fig1} \textbf{Optimised gate layout and dispersive sensing setup.} \textbf{(a)} False-coloured scanning electron micrograph of a nominally identical device. Cartoon cross sections (top, right) highlight the extended confinement gate C under the quantum dot accumulation gates $\gateone$, $\gatetwo$ that prevents the accumulation of electrons in the fan-out region. The confinement gate is extended to a thick region of oxide (right). Atomic layer deposited $\mathrm{Al_2 O_3}$ (blue layer) prevents shorts and reduces additional parasitic capacitance between the large resulting areas of overlapping aluminium (light blue) / thermal aluminium-oxide gate layers (red, orange, green, yellow). SRB, SLB and ST form the SET used to benchmark the gate-based readout. $\mathrm{G_T}$ controls the tunnel rate to the electron reservoir accumulated under R. \textbf{(b)} To detect charge tunnelling under the dot gate $\gatetwo$, we measure the phase shift of an $L-C$ resonant circuit via radio-frequency reflectometry. \textbf{(c)} Measured amplitude and phase response of the resonant circuit at millikelvin temperatures. \textbf{(d)} Charge stability diagram recorded using the SET showing a double dot formed under gates $\gateone$ and $\gatetwo$. Orange boxes indicate inter-dot charge transitions $(1,0)$-$(0,1)$ and $(4,0)$-$(3,1)$ investigated here.}
\end{figure}

The charge stability diagram of a double quantum dot defined under gates $\gateone$ and $\gatetwo$ is shown in Figure 1d. We study the dispersive response at the $(N_1,N_2) = (1,0)$ to $(0,1)$ inter-dot transition (Fig 2a). While the SET is sensitive to any change in local charge density, the gate-based sensing only detects charge tunnelling that can follow the dispersive sensing frequency $f_0$. The dot-to-reservoir transitions are not visible in the phase response due the slow (order kHz) tunnelling rates. At the location of the inter-dot charge transition, we observe a phase shift $\Delta \phi = 2.2$~mrad (Fig. 2a, right), caused by the added quantum capacitance of the electron tunnelling between the dots\cite{Petersson2010Charge,Cottet2011Mesoscopic,mizuta_quantum_2017} $C_q=(q_e \alpha_\epsilon^{G_2})^2\frac{d^2 E}{d \epsilon^2}$, where $\epsilon$ is the energy detuning $\epsilon = \mu_2-\mu_1$ between the dot chemical potentials, $\alpha_\epsilon^{G_2}$ is the lever-arm relating the voltage on $\gatetwo$ to detuning, $\alpha_\epsilon^{G_2}=\frac{d \epsilon}{dV_\mathrm{G_2}}$, and $E(\epsilon)= -\sqrt{(\epsilon/2)^2+t_c^2}$ the ground state energy dispersion for tunnel coupled dots. When the rf probe power is sufficiently low (Fig. 2b) the intrinsic width of the charge-transition is set by $t_c$, yielding $t_c=  12.0 \pm 1.5$~GHz (error margin here and elsewhere correspond to one standard deviation). From magneto-spectroscopy we find $\alpha_\epsilon^{G_2}= 0.10 \pm 0.03$~eV/V, which allows us to estimate an expected phase shift $\Delta \phi \approx \frac{\pi Q (q_e \alpha_\epsilon^{G_2})^2}{4 C_p t_c} = 0.5 - 2.6$~mrad, consistent with the measured response. The minimum integration time necessary to detect the phase shift is set by the effective noise temperature of the rf detection path and is limited by the noise temperature of the first amplifier. We extract a signal-to-noise ratio by comparing the change in the rf quadrature components due to the inter-dot tunnelling at $\epsilon = 0$ with the variance obtained from repeated sampling (Fig. 2c, see Methods for details). For an integration time $\taum  = 12$~ms, we can detect single electron tunnelling with a SNR of 2 (Fig. 2d).

\begin{figure}[tb]
	\centering
	\includegraphics[width=80mm]{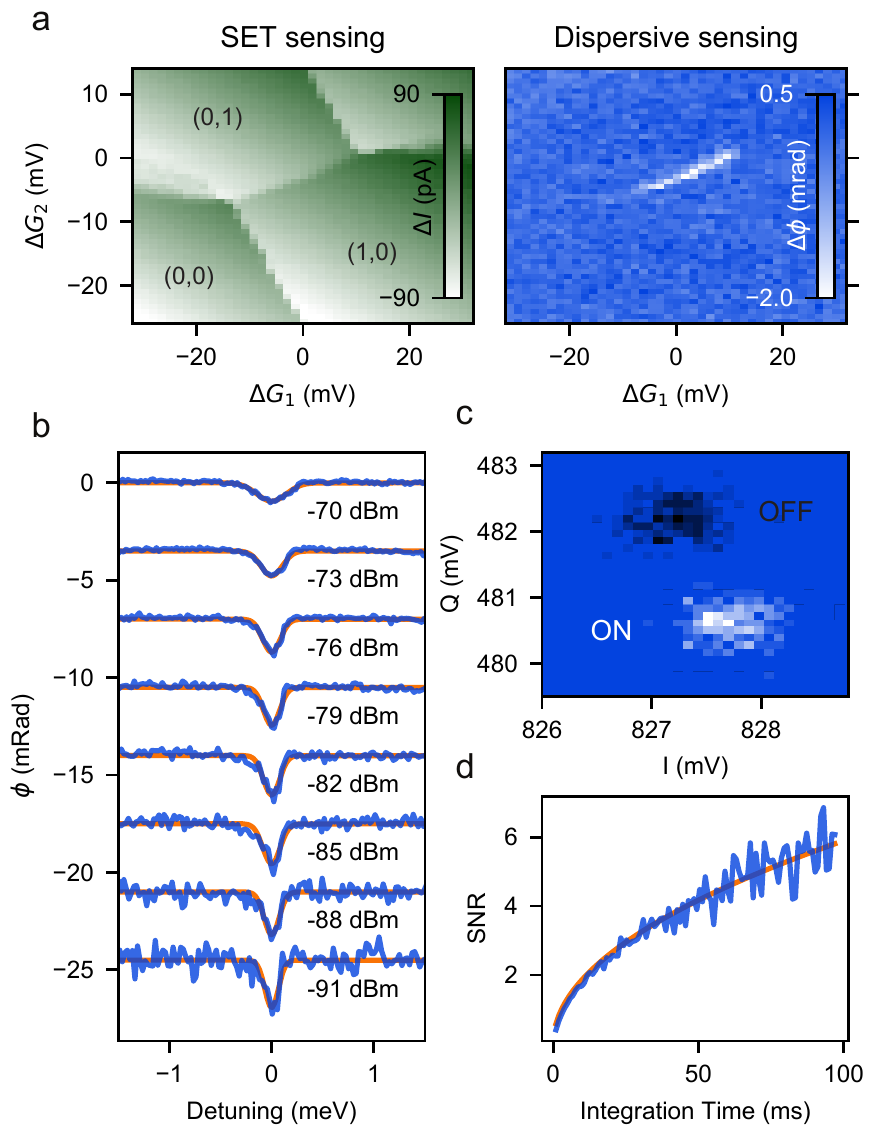}
	\caption{\label{fig:fig2} \textbf{Dispersive charge sensing of the double quantum dot.} \textbf{(a)} SET current $\Delta I$ (left) and rf phase response $\Delta \phi$ (right) obtained near the $(1,0)$-$(0,1)$ inter-dot transition. Current and phase relative to a reference point in the $(0,0)$ charge state is shown. \textbf{(b)} Phase response as a function of detuning $\epsilon$ for the inter-dot transition in (a), for varying rf probing powers. Estimated rf power at the device is shown. Gaussian fits to the data yield a constant phase response amplitude for probing powers up to  $-80$~dBm, after which the charge transition becomes power-broadened. \textbf{(c)} Using a power of $-83$~dBm, we obtain a histogram of the demodulated rf field quadrature components I and Q at the inter-dot transition (white, $\epsilon= 0$) and far detuned (black). The histogram shown is obtained by repeated sampling of the I and Q components with an analog bandwidth of $100$~kHz, averaging for an integration time of $12$~ms. 
	\textbf{(d)} SNR obtained from histograms in (c), as a function of integration time for this probing power.}
\end{figure}

Having established and characterised the gate-based detection of single electron tunnelling, we now focus on the spin-readout capability. When a double quantum dot is occupied by two electrons, the Pauli exclusion principle prevents inter-dot tunnelling for all but the singlet spin state. This provides a means to probe the spin configuration via the change in gate-impedance near the inter-dot charge transition. In silicon, the valley degree of freedom acts as a low-lying orbital state that can break the spin blockade, as triplets can populate the excited valley state in a doubly occupied dot. Here we operate at the $(4,0)$-$(3,1)$ inter-dot charge transition to avoid the low-lying valley state (estimated valley splitting $50$~µeV
) which would otherwise serve to lift the spin blockade. At the $(4,0)$-$(3,1)$ transition, two electrons in a spin-singlet state always fill the lower valley of dot 1, and do not affect the tunnel dynamics of the remaining two electrons. In Figure 3c, we compare the SET current and dispersive response for two different state initialisation protocols\cite{Fogarty2017Integrated} ($A1$/$A2$ and $B1$/$B2$, Fig. 3a,b). For protocol $A$, plunging into $(4,0)$ from $(3,0)$ ($A1 \rightarrow A2$) initialises a spin-singlet state due to the large energy gap to the first excited triplet state. The singlet initialisation is confirmed by the detection of a dispersively detected phase shift near the inter-dot transition (See Supplementary Figure S2).
Conversely, for protocol $B$, at low external magnetic field (for the data in Fig. 3c and Fig. 4 we set $B_\mathrm{ext}=250$~mT $\ll \gamma_e T_e/\mu_B$, with $\gamma_e$ the electron gyromagnetic ratio and $T_e$ the electron temperature of the reservoir), unloading an electron by pulsing into $(3,1)$ from $(4,1)$ ($B1 \rightarrow B2$),  randomises the spin state into a mixture of singlet and triplets. The resulting partial blockade of tunnelling can be observed when subtracting the signals obtained from initialisation protocols $A$ and $B$ (Fig. 3c), and shows the expected Pauli spin blockade triangle (cut off by the first available triplet-like state of dot 1 at $2.0 \pm 0.3$~meV, attributed to the first orbital excited state). We further verify the spin nature of the blockade by probing the coupling between the hybridised singlet state $S$ and the lowest triplet spin state $\Tm$~(Fig. 3b). The $S-\Tm$ anti-crossing was previously studied in this system\cite{Fogarty2017Integrated,Tanttu2018Controlling}, and is mediated here by a combination of hyperfine and spin-orbit interaction. After $S$-initialisation (protocol $A$), we plunge into $(3,1)$ along the detuning axis (3) shown in Fig. 3a and vary plunge depth $\epsilon$ and external magnetic field $B_\mathrm{ext}$. When $\epsilon$ corresponds to the location of the $S-\Tm$ anti-crossing, we expect a reduced singlet return probability, resulting in Pauli spin blockade signal at the readout point (4,RO). $I$ and $\phi$ recorded at a reference point (5,Ref) in $(3,0)$ are subtracted to counter slow drifts. We observe a characteristic spin-funnel\cite{Petta2005Coherent} via both SET and phase response, mapping out location of the $S-\Tm$ anti-crossing (Fig. 3d). Based on a Hamiltonian model\cite{Fogarty2017Integrated}, the shape of the funnel determines the tunnel coupling at this inter-dot transition, $t_c^{(4,0)-(3,1)}=39.5 \pm 2$~GHz, consistent with the observed phase-shift.

\begin{figure}[hbt]
	\centering
	\includegraphics[width=80mm]{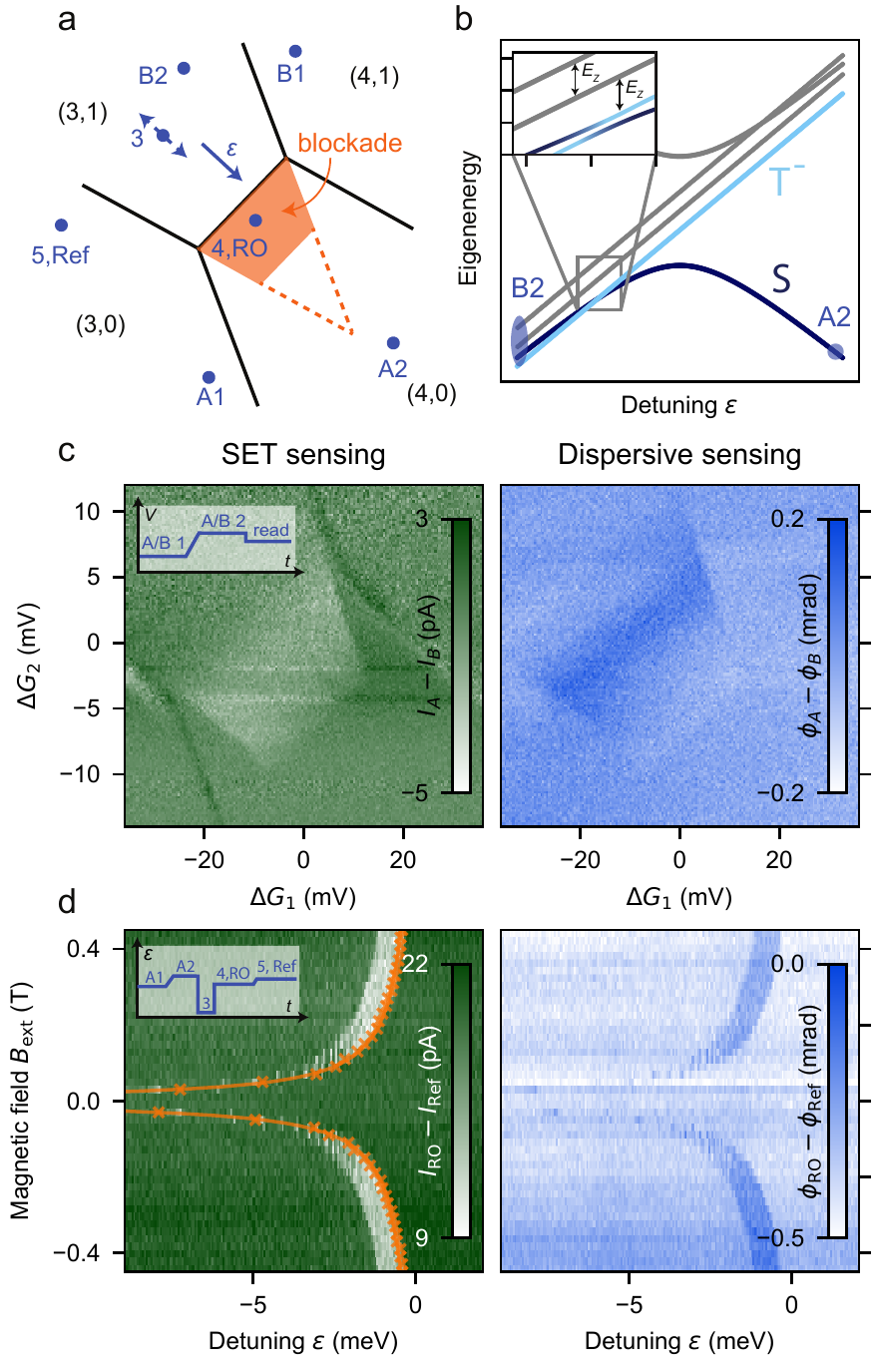}
	\caption{\label{fig:fig3} \textbf{Dispersive spin blockade readout.} \textbf{(a)} Schematic charge stability near the $(4,0)$-$(3,1)$ inter-dot charge transition. The orange region shows the expected Pauli spin blockade triangle, truncated by an excited $(4,0)$ triplet state. Blue points indicate gate voltages used in pulse sequences for (c) and (d), blue arrow shows the axis of dot chemical potential detuning $\epsilon$. \textbf{(b)} Energy diagram for the five lowest eigenstates near $(4,0)$-$(3,1)$, as a function of $\epsilon$. Hybridised spin singlet state $S$, and lower polarised spin triplet state $\Tm$~are indicated.  Also indicated are the approximate initialisation points $A2$, $B2$. (inset) $S-\Tm$ anti-crossing at finite $\epsilon$ as a function of Zeeman splitting $E_Z=g \mu_B B_\mathrm{ext}$. \textbf{(c)} SET current (left) and dispersive response (right) measured after initialising either via $A1 \rightarrow A2$, to initialise $S$, or via $B1 \rightarrow B2$ to initialise a mixed state between $S$ and the three triplets (pulse sequence is shown in inset). The difference between protocols $A$ and $B$ is shown, highlighting a spin blockade region where triplet states are prevented from tunnelling. \textbf{(d)}  A characteristic spin-funnel is observed, mapping out the $S-\Tm$ anti-crossing as a function of Zeeman splitting $E_Z$, confirming the spin nature of the blockade. A fit to the data (orange line) yields the tunnel coupling for this inter-dot transition. Used pulse-sequence ($A1$-$A2$-3-4,RO-5,Ref) is shown in the (inset), see Main text for details.}
\end{figure}

To estimate the spin-readout signal one can obtain in a single shot (for a single preparation of the qubit), we make a histogram of single-shot experiments (Fig. 4a), where a slow pulse to the $S-\Tm$ crossing is used to initialise an evenly mixed $S+\Tm$ state (for reference a $S$-initialised state, using only initialisation protocol $A$, is also shown in grey in the marginal distributions, Fig. 4a, top and right). The single-shot measurements show a bi-modal distribution, with a clear correlation between the SET current and dispersively detected phase-shift. This allows us to estimate the $S-\Tm$ spin readout fidelity, using a model\cite{Barthel2009Rapid} for singlet-triplet readout to fit the marginal distributions for the $S$ and $S+\Tm$ initialised states (see Methods for details). The model takes into account the finite triplet blockade lifetime $\Tone = 4.5 \pm 0.5$~ms and the $\Tm$ initialisation fraction ($p_\Tm = 0.50 \pm 0.03$), determined from an independent measurement presented in Figure 4b. We first fit the SET and dispersive readout distributions independently (Fig. 4a, top and right, solid lines show the resulting model output). Performing the fit for various integration times $\tau_m$~ and choosing an optimal threshold yields the average $S$ - $\Tm$ readout fidelity $F_\mathrm{avg} = F_S/2 + F_{\Tm}/2$. We find $F_\mathrm{avg}^\mathrm{SET}=88.2\pm1.9$~\% for an optimal $\tau_m^\mathrm{SET} = 1.0$~ms. Similarly, for the dispersive readout, $F_\mathrm{avg}^\mathrm{dispersive}=74.5\pm1.9$~\% for $\tau_m^\mathrm{dispersive}=2.6$~ms. Then, as a cross-check, we use the SET readout to bin single shot events into clear $S$ or $\Tm$ outcomes (Methods). From the overlap of the resulting dispersive readout histograms (Fig. 4c), and again taking into account $T_1$, we obtain $F_\mathrm{avg}^\mathrm{dispersive}=73.3\pm1.2$~\% for $\tau_m^\mathrm{dispersive}=2.0$~ms. We note that the Pauli spin blockade based readout used here is directly applicable to realise local parity measurements for error detection in arrays of single-spin based qubits\cite{Veldhorst2017Silicon,Jones2018Logical}.

\begin{figure}[h!bt]
	\centering
	\includegraphics[width=80mm]{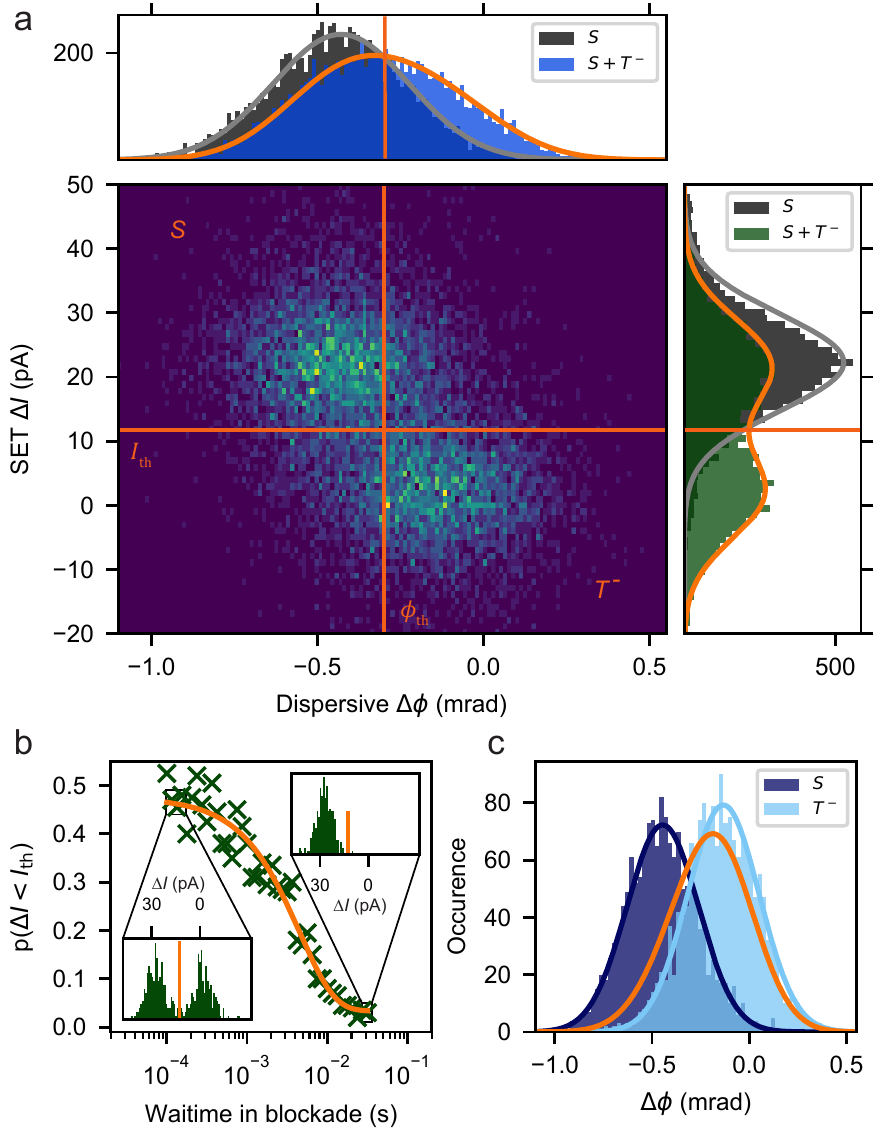}
	\caption{\label{fig:fig4} \textbf{Gate-based single-shot spin readout characterisation.} \textbf{(a)} Using the $S-\Tm$ anti-crossing to initialise a mixture of $S$ and $\Tm$ (see pulse sequence in Fig. 3d), we correlate the SET-current change ($\Delta I=I_\mathrm{RO}-I_\mathrm{Ref}$) and phase shift ($\Delta \phi= \phi_\mathrm{RO}-\phi_\mathrm{Ref}$). Shown is a histogram of 10000 single shots with an integration time $\tau_m^\mathrm{SET} = 1.0$~ms and $\tau_m^\mathrm{dispersive} = 2.5$~ms. We find a bi-modal distribution, with one peak corresponding to a $S$ outcome, and the other to a $\Tm$ outcome. As expected the SET outcome correlates with the dispersive one. Marginal distributions for $\Delta I$ and $\Delta \phi$ are shown on (right), (top), respectively. Also shown in grey is the histogram obtained for an identical readout after $S$-initialisation, using protocol A. Solid lines are fits to a model for singlet-triplet readout\cite{Barthel2009Rapid}, yielding an estimate for the single-shot readout fidelity (see Main text). \textbf{(b)} The optimal integration time is limited by the blockaded state lifetime $T_1$. To characterise $T_1$, we repeat the experiment with a varying wait-time at the readout point (4,RO), before starting to integrate the data. The same continuous rf probing power ($-61$~dBm) is used for (a-c). For short wait-times the SET current distribution reflects the initial state (top left inset), whereas for very long wait-times the blockaded state has fully decayed (bottom right inset). Fitting an exponential decay, we find a blockade lifetime of $\Tone = 4.5 \pm 0.5$~ms. \textbf{(c)} Using the SET readout results in (a) to bin single shots into $S$ or $\Tm$ outcomes, we can separately determine the dispersive response for $S$ and $\Tm$ states. Gaussian fits (solid blue lines) to the resulting histograms and modelling the effect of $T_1$ (orange line), yields an accurate measurement for the spin readout fidelity (see Main text).}
\end{figure}

Although here the SET outperforms the dispersive gate-based sensing, modest improvements to the resonator circuit should significantly boost the gate-based readout fidelity. For example, superconducting spiral inductor-based resonators have been shown to attain quality factors exceeding 100, with typical parasitic capacitance values around 0.3~pF\cite{Colless2013Dispersive,Hornibrook2014Frequency}.  This would increase the observed phase shift eight-fold, resulting in an average gate-based spin readout fidelity exceeding 99\% (Methods). Further improvements can be obtained using optimised external matching techniques\cite{ahmed_radio-frequency_2018} or parametric amplification. Moreover, we emphasise that even if an on-chip electrometer outperforms a gate-based solution in terms of readout sensitivity, its benefits must be weighed against the added complexity as the number of qubits is increased. Finally, the gate-based sensing naturally allows frequency multiplexing by operating resonant circuits with varying inductor values on each gate\cite{Hornibrook2014Frequency}. 

In summary, we have characterised a gate-based approach for spin-qubit measurements in a future silicon quantum processor.
The signal-to-noise ratio obtained with a simple resonant circuit is sufficient to read out the electronic spin state in a single shot. Our results, together with contemporaneous results in several other silicon quantum dot architectures\cite{Vandersypen2018Results,pakkiam_single-shot_2018,urdampilleta_gate-based_2018}, open a path to the readout of many spin qubits in parallel, using a compact gate design that will be needed for large-scale quantum processors of the future.

\clearpage
\subsection{Methods}
\subsubsection{Experimental methods}
The device was fabricated on natural silicon using multi-level gate-stack MOS technology\cite{Veldhorst2014addressable}. Four layers of gates with thickness 25, 45, 65, and 65 nm were fabricated on top of 5.9 nm thick thermally grown $\mathrm{Si O_2}$ using electron beam lithography and aluminium evaporation and separated by thermally grown aluminium-oxide. After the first layer was completed, we performed 50 cycles of atomic layer deposition at 275°C (intended thickness 6 nm), to grow an additional insulating film of $\mathrm{Al_2 O_3}$ between gate layers 1 and 2 (see Figure 1a). 
The device was bonded to a printed circuit board (PCB) in a copper enclosure and cooled down in a dilution refrigerator with electron base temperature of 180 mK. Ceramic chip inductors (Coilcraft 1206CS-471 and 1206CS-122) and capacitor were surface mounted on the PCB holding the device, to provide the resonant circuit and DC bias (see Figure 1b). rf reflectometry measurements were performed employing either a vector network analyser (Keysight Fieldfox, for the data in Fig. 1c) or a demodulation setup using a rf source (Stanford Research Systems SG380) and IQ demodulator (Polyphase AD0105). The rf detection path consisted of a directional coupler (Mini-Circuits ZEDC-15-2B), cryogenic amplifier (Miteq AFS3-00100200-10-CR-4) mounted to the 4K plate and two room temperature amplifiers (Mini-Circuits ZFL-1000LN+). The demodulated signal was further amplified and filtered at 100 kHz, 48 dB/octave (Stanford Research Systems SIM910 and SIM965) before digitisation at 200 kS/s (Gage Octopus CS8389). Charge stability diagrams were obtained using a double lock-in technique with dynamic voltage compensation.
In order to estimate the signal-to-noise ratio for the detection of inter-dot tunnelling, we repeatedly sample the reflected rf quadratures at $\epsilon=0$, ($I_\mathrm{ON},Q_\mathrm{ON}$) and far-detuned from the interdot transition ($I_\mathrm{OFF},Q_\mathrm{OFF}$, keeping the same voltage on $\gatetwo$ in order to avoid effects from changes in $C_p$, see Supplementary Figure S1). We define the displacement signal power $S= \langle I_\mathrm{ON}-I_\mathrm{OFF} \rangle^2 + \langle Q_\mathrm{ON}-Q_\mathrm{OFF} \rangle^2$, where $\langle\cdot\rangle$ denotes the mean over the repeated samples. Similarly, we define the noise power $N = \mathrm{std}[(I_\mathrm{ON}-I_\mathrm{OFF})^2 + (Q_\mathrm{ON}-Q_\mathrm{OFF})^2]$, where $\mathrm{std}[\cdot]$ denotes the standard deviation of the repeated samples. The SNR is $S/N$. 
\subsubsection{Single-shot readout fidelity analysis}
Using the model by Barthel et al.\cite{Barthel2009Rapid}, we describe the single-shot readout histograms for the SET and dispersive response by $N(X) = N_\mathrm{tot}\left[(1-p_\Tm)n_S(X) + p_\Tm n_\Tm (X)\right]X_\mathrm{binsize}$. For the SET readout, we replace $X$ by the SET current $\Delta I$ in above equation, while for dispersive readout we replace $X$ by phase shift $\Delta\phi$. $p_\Tm$ is the $\Tm$ initialisation fraction, $N_\mathrm{tot}=10000$ is the total number of single shot readout events, $X_\mathrm{binsize}$ is the histogram bin-size, $n_S(X) = \frac{1}{\sqrt[]{2 \pi \sigma}} \exp \left[ -\frac{(X-X_S)^2}{2 \sigma^2} \right]$, and $n_\Tm(X) = \frac{1}{\sqrt[]{2 \pi \sigma}} \exp \left[-\frac{\taum}{\Tone} \right] \exp \left[ -\frac{(X-X_\Tm)^2}{2 \sigma^2} \right] + \int_{X_S}^{X_\Tm} \frac{1}{\sqrt[]{2 \pi \sigma}} \frac{\taum}{\Tone} \frac{1}{\left| X_\Tm - X_S \right|} \exp \left[-\frac{\taum}{\Tone} \frac{(\tilde{X}-X_S)}{(X_\Tm - X_S)} \right] \times \exp \left[ -\frac{(X-\tilde{X})^2}{2 \sigma^2} \right] d\tilde{X}$, with $\tau_m$ the integration time and $T_1$ the blockade lifetime. To fit the model to the measured histograms, we proceed as follows: first we use the singlet initialised data (grey histograms in Fig. 4a) to determine $X_S$ (fixing $p_\Tm=0$). Then, the $S$ + $\Tm$ initialised histogram is fit to obtain $X_\Tm$ and $\sigma$, where $\Tone  = 4.5$~ms and $p_\Tm=0.5$ are fixed to the values obtained from the data in Figure 4b. $X_S$, $X_\Tm$ and $\sigma$ are independently fit for the SET and dispersive readout. This procedure is repeated for increasing integration time $\tau_m$, and for each integration time the optimal threshold $X_\mathrm{th}$ is chosen that maximises the average readout fidelity $F_\mathrm{avg} = F_S/2 + F_\Tm/2$, with $F_S=1-\int_{X_\mathrm{th}}^\infty  n_S(\tilde{X}) d\tilde{X}$, $F_\Tm = 1-\int_{-\infty}^{X_\mathrm{th}} n_\Tm(\tilde{X}) d\tilde{X}$ (for the case $X_S > X_\Tm$, the integration boundaries are inverted). The resulting readout fidelities $F_\mathrm{avg}(\taum)$ are shown in Supplementary Figure S3 (green and blue lines). The data in Figure 4c is obtained by using the SET signal to categorise single-shot events that have a clear $S$ or $\Tm$ result: we categorise single-shot readout events that have $\Delta I < I_\mathrm{th}^\Tm$ as $\Tm$ readout events, and $\Delta I > I_\mathrm{th}^S$ as $S$ readout events. Here, $I_\mathrm{th}^\Tm = I_\mathrm{th}-12$pA, ~$I_\mathrm{th}^S = I_\mathrm{th} +12$~pA, where $I_\mathrm{th} = 12.7$~pA is the optimal single-shot readout threshold for the SET at time $\tau_m^\mathrm{SET}=1$~ms. We fit a single Gaussian to each of the $S$ and $\Tm$ categorised event histograms (varying $\tau_m^\mathrm{dispersive}$) to obtain an accurate value for $\Delta \phi_S$ and $\Delta \phi_\Tm$ respectively, as well as $\sigma = \sigma_S/2 + \sigma_\Tm/2$. Inserting these values (and $T_1$) in the model above is sufficient to calculate $F_\mathrm{avg}^\mathrm{dispersive}$,  choosing the optimal $\phi_\mathrm{th}$ for each $\tau_m^\mathrm{dispersive}$ (orange line in Supplementary Figure S3).

To estimate the readout fidelity that could be obtained for improved parasitic capacitance $C_p$ and resonator quality factor $Q$, we use $\Delta \phi \sim \frac{Q}{C_p}$, which for the values given in the main text gives a factor 7.8 increase in $\Delta \phi$ compared to current values. Setting $\phi_S \rightarrow 7.8\phi_S$ in the model above (assuming $\sigma$ unchanged) gives $F_\mathrm{avg}^\mathrm{dispersive} = 0.996$.

\subsection{Acknowledgements}
We thank M. House and A. Laucht for helpful discussions and C. Escott for valuable feedback on the manuscript. We acknowledge support from the US Army Research Office (W911NF-17-1-0198), the Australian Research Council (CE11E0001017), and the NSW Node of the Australian National Fabrication Facility. The views and conclusions contained in this document are those of the authors and should not be interpreted as representing the official policies, either expressed or implied, of the Army Research Office or the U.S. Government. The U.S. Government is authorised to reproduce and distribute reprints for Government purposes notwithstanding any copyright notation herein. M.F.G.Z. is supported by the Horizon 2020 programme through Grant Agreement No. 688539. B.H. acknowledges support from the Netherlands Organisation for Scientific Research (NWO) through a Rubicon Grant. This work was partly supported by the Winton Fund for the Physics of Sustainability.
\subsection{Author contributions}
A.W., B.H. and A.J. performed the experiments. A.W. designed the device with input from A.R. and M.F.G., A.W. and F.E.H. fabricated the device with A.S.D’s supervision. T.T., C.H.Y. and A.M. contributed to the preparation of experiments and experimental systems. A.R. and M.F.G supervised early experiments. A.W., B.H. and A.J. designed the experiments under the supervision of A.S.D., with D.R. contributing to results discussion and interpretation. B.H. and A.W. wrote the manuscript with input
from all co-authors.

\bibliographystyle{naturemag}
\bibliography{ref-extracts.bib}
\end{document}


\title{Gate-based single-shot readout of spins in silicon - SUPPLEMENTARY INFORMATION}
\author{A.~West}
\thanks{These authors contributed equally}
\author{B.~Hensen}
\thanks{These authors contributed equally}
\affiliation{Centre for Quantum Computation and Communication Technology,
School of Electrical Engineering and Telecommunications,
The University of New South Wales, Sydney, New South Wales 2052, Australia}
\author{A.~Jouan}
\affiliation{ARC Centre of Excellence for Engineered Quantum Systems, School of Physics, 
The University of Sydney, Sydney, NSW 2006, Australia}
\author{T.~Tanttu}
\author{H.~Yang}
\affiliation{Centre for Quantum Computation and Communication Technology,
School of Electrical Engineering and Telecommunications,
The University of New South Wales, Sydney, New South Wales 2052, Australia}
\author{A.~Rossi}
\affiliation{Cavendish Laboratory, University of Cambridge, J.J. Thomson Avenue, Cambridge CB3 0HE, United Kingdom}
\author{M.F.~Gonzalez-Zalba}
\affiliation{Hitachi Cambridge Laboratory, 
J.J. Thomson Avenue, Cambridge CB3 0HE, United Kingdom}
\author{F.E.~Hudson}
\author{A.~Morello}
\affiliation{Centre for Quantum Computation and Communication Technology,
School of Electrical Engineering and Telecommunications,
The University of New South Wales, Sydney, New South Wales 2052, Australia}
\author{D.J.~Reilly}
\affiliation{ARC Centre of Excellence for Engineered Quantum Systems, School of Physics, 
The University of Sydney, Sydney, NSW 2006, Australia}
\affiliation{Microsoft Corporation, Station Q Sydney, The University of Sydney, Sydney, NSW 2006, Australia}
\author{A.S.~Dzurak}
\email{b.hensen@unsw.edu.au, a.dzurak@unsw.edu.au}
\affiliation{Centre for Quantum Computation and Communication Technology,
School of Electrical Engineering and Telecommunications,
The University of New South Wales, Sydney, New South Wales 2052, Australia}

\maketitle


\begin{figure}[htb]
	\centering
	\includegraphics[width=89mm]{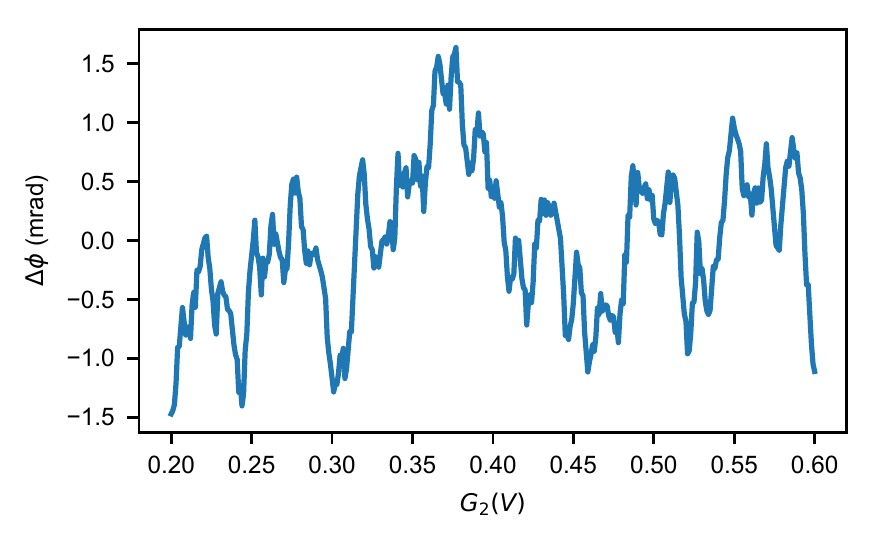}
	\caption{\label{fig:figS1} \textbf{Residual $\gatetwo$ gate voltage dependent phase shift} Here we characterise the residual phase shift for large variations of the voltage on $\gatetwo$. We measure $\phi$ for a $\gatetwo$ voltage range spanning multiple charging voltages, with the quantum dot $\gateone$ fully depleted. We find a reproducible pattern with a peak-to-peak phase shift of a few mrad. We attribute this phase shift to residual coupling of the $\gatetwo$ electrode to charge in the fan-out region. A phase shift due to the voltage on $\gateone$ is not observed.}
\end{figure}

\begin{figure}[htb]
	\centering
	\includegraphics[width=89mm]{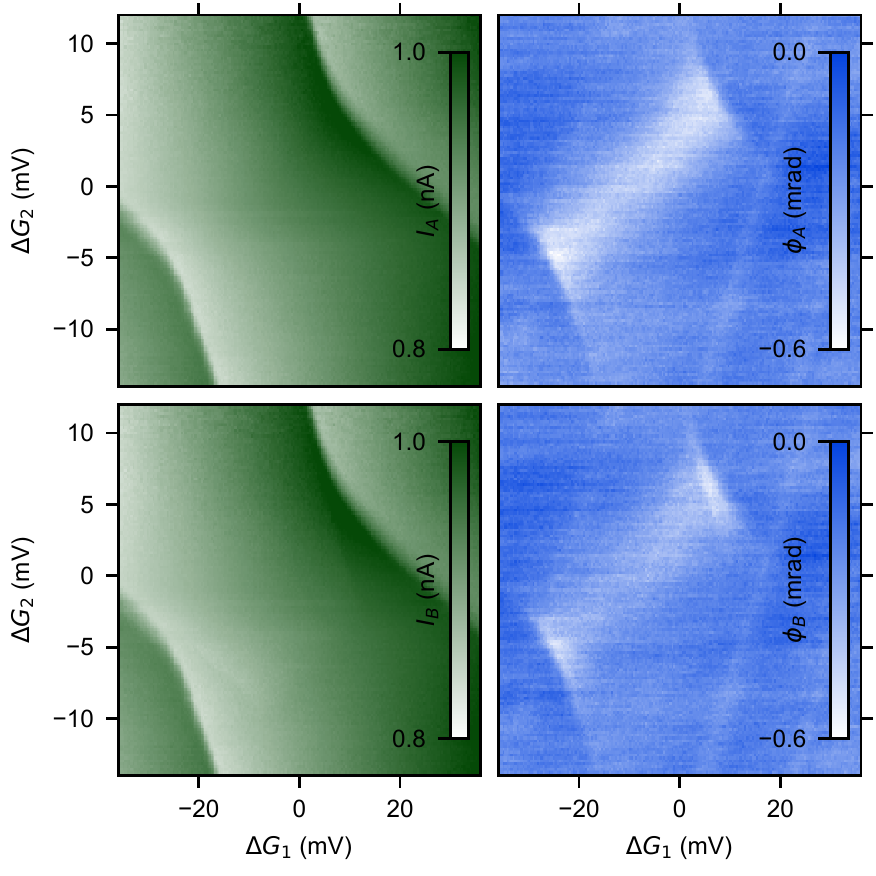}
	\caption{\label{fig:figS2} \textbf{Spin-blockade data for initialisation protocols A, B before subtraction} \textbf{(a)} SET current (left) and dispersive phase (right) for initialisation protocol A (see Main text), expected to initialise a spin-singlet state. \textbf{(b)} same for protocol B, expected to initialise a mixture of singlet and triplet spin states. In (b), the dispersive tunnelling response is reduced due to the blockaded triplet states. The blockade region can also be observed for the SET current. The data is recorded one horizontal line after another. For the SET current we subtracted the mean and divided by the variance of each horizontal line, adding the mean and multiplying by the variance of the whole data-set to correct for slow SET sensitivity drifts. For the dispersive phase we similarly subtract the mean of each line and add the mean of the whole data-set. In Fig. 3c of the Main text we only subtract the mean of each line.}
\end{figure}

\begin{figure}[htb]
	\centering
	\includegraphics[width=89mm]{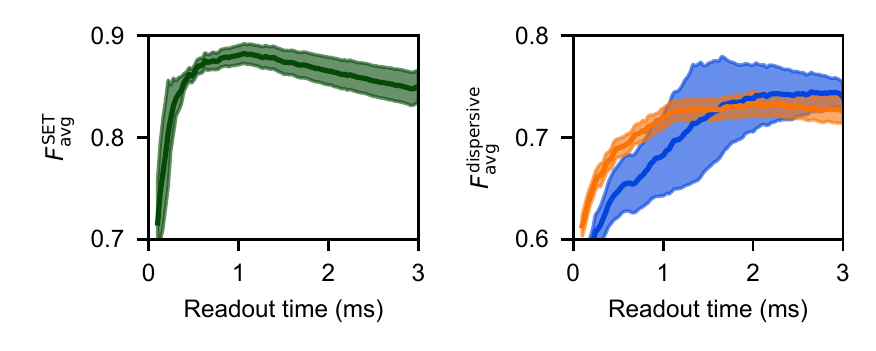}
	\caption{\label{fig:figS3} \textbf{Average $S-\Tm$ readout fidelity versus integration time}. (left) SET readout, green line and (right) dispersive readout, blue line are based on fitting the model for singlet triplet readout (see Methods) to the histograms for the SET and dispersive data independently. The orange line (right) is obtained from binning each single shot into $S$ or $\Tm$ events using the SET signal and using the resulting conditional histograms as input to the model. See Methods for details.}
\end{figure}